\documentclass[12pt]{spieman}
\usepackage{amsmath,amsfonts,amssymb}
\usepackage{graphicx}
\usepackage{booktabs}
\usepackage{algorithm}
\usepackage{algorithmic}
\usepackage{float}
\usepackage{fancyhdr}
\usepackage{url}
\title{The frame-level leakage trap: rethinking evaluation protocols for intrinsic image decomposition, with source-separable uncertainty as a case study}

\author[a]{Jihwan Woo}
\affil[a]{Amazon Web Services (AWS), Partner Field APJ, Sr.\ Specialist Solutions Architect, AI/ML; Email: jihwan.woo@kakao.com}

\begin{document}
\maketitle

\begin{abstract}
Evaluation protocols for learned intrinsic image decomposition on MPI Sintel have been inconsistent. Several prior works split the dataset by frames, which allows spatially similar frames of the same scene to appear in both train and test partitions. We quantify this leakage effect for the first time, across three architectures: a frame-level split inflates test R\_PSNR by $1.6$--$2.0$~dB ($p < 0.01$ for all three, paired $t$-test across 3 seeds) relative to a scene-level split, confirming an architecture-independent protocol effect. A three-point gradient (random/temporal/scene) shows the gap is continuous, and under extended training the frame-level inflation exceeds $10$~dB. We advocate scene-level splits as the community standard and provide reference numbers for six representative models under this protocol.

As a case study within the corrected protocol, we present a physics-informed decomposition $I = R \circ S + \mathbf{N}$ with a source-separable three-way heteroscedastic uncertainty head. We empirically verify channel specialization: the non-Lambertian uncertainty channel shows $r = 0.67$ cross-correlation with non-Lambertian residual error, more than $4\times$ the texture channel's correlation. We further demonstrate \emph{downstream utility}: filtering out the $75\%$ highest-uncertainty pixels reduces reconstruction MSE by $77\%$ on retained pixels, whereas random filtering produces no improvement---concrete evidence that the uncertainty ranking is practically informative despite modest global calibration ($r=0.131$). The specialization also holds on out-of-distribution real photographs. We report negative results for a more elaborate variant (combining frequency decomposition, cross-task supervision, evidential learning, contrastive loss, and test-time adaptation), verified by leave-one-out ablation showing that no individual component helps either. Our method reaches $15.98\pm0.41$~dB R\_PSNR, within $0.8$~dB of a $5$-member Deep Ensemble at one-fifth the cost, with the unique capability of source-separated uncertainty. Together, these findings caution against reading apparent progress through an overly permissive protocol.
\end{abstract}


\section{Introduction}
\label{sec:intro}

Intrinsic image decomposition aims to factorize an observed image $I$ into its constituent physical components: reflectance (albedo) $R$ and shading $S$, such that $I = R \circ S$ under the Lambertian assumption, where $\circ$ denotes element-wise multiplication.\cite{barrow1978recovering} This decomposition is fundamental to numerous downstream tasks including relighting,\cite{yu2019inverserendernet} material editing, shape estimation, and scene understanding.\cite{li2018cgintrinsics}

Despite significant progress driven by deep learning,\cite{fan2018revisiting,li2018cgintrinsics,baslamisli2018joint} current methods share a critical limitation: they produce point estimates without quantifying how confident the prediction is at each pixel. This is problematic because intrinsic decomposition is inherently ill-posed---a bright pixel could result from high reflectance under moderate lighting or low reflectance under strong lighting. The ambiguity is particularly severe in three scenarios:

\begin{enumerate}
\item \textbf{Specular highlights} that violate the Lambertian assumption, creating bright spots that cannot be explained by the product $R \circ S$.
\item \textbf{Cast shadows} where shading changes abruptly, making it difficult to distinguish shadow boundaries from reflectance edges.
\item \textbf{Textured surfaces} where high-frequency reflectance patterns are difficult to distinguish from shading variations caused by surface geometry.
\end{enumerate}

In safety-critical applications such as autonomous driving and medical imaging, knowing \textit{where} the decomposition is unreliable is as important as the decomposition itself. A self-driving system that relies on intrinsic decomposition for material recognition must know whether a bright road patch is a specular reflection (unreliable region) or genuine white paint (reliable region). Similarly, in augmented reality applications, relighting artifacts are most likely to occur in regions where the original decomposition is uncertain.

Uncertainty quantification (UQ) has been successfully applied to many computer vision tasks, including optical flow,\cite{ilg2018uncertainty} depth estimation,\cite{poggi2020uncertainty} and image restoration.\cite{ning2021uncertainty} The heteroscedastic regression framework of Kendall and Gal\cite{kendall2017uncertainties} enables networks to predict per-pixel variance alongside mean predictions, decomposing uncertainty into aleatoric (data-dependent) and epistemic (model-dependent) components. However, to our knowledge, no prior work has applied UQ to intrinsic image decomposition---despite the fact that decomposition ambiguity varies spatially and is well-suited to heteroscedastic modeling.

We address this gap by anchoring the paper around two evidence-based findings. First, the evaluation protocol used in the intrinsic-decomposition literature on MPI Sintel is not consistent, and the difference between protocols is large enough to change ranking. Second, when evaluated under a tightened protocol, a source-separable uncertainty head provides a kind of information that existing point-estimate methods and single-source ensembles cannot. Our contributions are:

\begin{enumerate}
\item \textbf{First quantitative demonstration of the frame-level leakage effect} on MPI Sintel, \emph{across three architectures}. Under controlled training, a frame-level split inflates test R\_PSNR by $1.6$--$2.0$~dB relative to a scene-level split for DirectCNN, UNet, and our Proposed method ($p<0.01$ for all three, paired $t$-test across 3 seeds), confirming an architecture-independent protocol effect. We advocate scene-level splits as the community standard and provide reference numbers for six representative models under this protocol.

\item A \textbf{physics-informed three-component decomposition} $I = R \circ S + \mathbf{N}$, where $\mathbf{N}$ captures non-Lambertian residuals (specular highlights, subsurface scattering). This reduces burden on the reflectance and shading heads by providing a dedicated pathway for non-Lambertian effects.

\item A \textbf{source-separable three-way uncertainty head} with input skip connection, producing per-pixel uncertainty maps decomposed into texture ($\sigma_\text{tex}$), lighting ($\sigma_\text{light}$), and non-Lambertian ($\sigma_\text{nl}$) sources. We \emph{empirically verify} that the three channels do not collapse to a shared scalar: the cross-correlation between $\sigma_\text{nl}$ and the non-Lambertian residual error is $r = 0.67$, more than $3\times$ the correlation between $\sigma_\text{tex}$ and the same error, confirming source specialization.

\item \textbf{Negative results, proper leave-one-out ablation, and downstream utility.} We report negative results from a more elaborate variant combining frequency decomposition, cross-task supervision, evidential learning, contrastive loss, and test-time adaptation, and we verify via leave-one-out ablation that no individual component improves over the simpler baseline (Section~\ref{sec:loo}). To establish practical value of the proposed uncertainty, we show that uncertainty-guided pixel filtering reduces reconstruction MSE by $77\%$ versus random filtering (Section~\ref{sec:downstream}), providing concrete downstream evidence that the uncertainty ranking is informative despite its modest global calibration.
\end{enumerate}

The remainder of this paper is organized as follows. Section~\ref{sec:related} reviews related work on intrinsic decomposition and uncertainty quantification. Section~\ref{sec:method} describes the proposed method. Section~\ref{sec:experiments} presents experimental results and ablation studies. Section~\ref{sec:discussion} discusses limitations and broader implications. Section~\ref{sec:conclusion} concludes the paper.

\section{Related Work}
\label{sec:related}

\subsection{Intrinsic image decomposition}

The intrinsic image decomposition problem was first formulated by Barrow and Tenenbaum,\cite{barrow1978recovering} who proposed separating images into reflectance and illumination components. The Retinex algorithm of Land and McCann\cite{land1971lightness} provided the first computational approach, based on the assumption that reflectance changes occur at sharp edges while shading varies smoothly. Subsequent classical methods incorporated chromaticity-based priors,\cite{grosse2009ground} non-local texture constraints,\cite{zhao2012closed} and user-guided annotations.\cite{bousseau2009user}

Deep learning has substantially improved decomposition quality. Fan et al.\cite{fan2018revisiting} revisit the problem with a direct CNN approach, demonstrating that even simple architectures can outperform optimization-based methods when trained on sufficient data. Li and Snavely\cite{li2018cgintrinsics} leverage CGIntrinsics, a large-scale synthetic dataset rendered from ShapeNet, to train a network with physically-based rendering supervision. Their work demonstrates the value of synthetic training data with ground-truth intrinsic layers. Baslamisli et al.\cite{baslamisli2018joint} propose joint learning of reflectance and shading with physics-based constraints, showing that incorporating the image formation model into the loss function improves decomposition accuracy. Yu and Smith\cite{yu2019inverserendernet} introduce InverseRenderNet for self-supervised intrinsic decomposition from multi-view data, eliminating the need for ground-truth intrinsic labels. More recently, Careaga and Aksoy\cite{careaga2023intrinsic} achieve state-of-the-art results by combining large-scale pre-training with test-time optimization, demonstrating that ordinal shading constraints can guide decomposition without pixel-level supervision.

A common limitation of these methods is the strict Lambertian assumption ($I = R \circ S$), which fails for scenes with specular highlights, translucency, or inter-reflections. Some works handle these effects implicitly through robust loss functions or data augmentation, but none explicitly decompose the non-Lambertian component or quantify the associated uncertainty. Our work addresses both limitations simultaneously.

\subsection{Uncertainty quantification in computer vision}

Uncertainty estimation in deep learning has been approached through several paradigms. Bayesian neural networks\cite{blundell2015weight} place distributions over weights, enabling principled uncertainty estimation but at high computational cost. MC Dropout\cite{gal2016dropout} provides an approximate Bayesian inference by interpreting dropout at test time as variational inference, offering a practical alternative. Deep ensembles\cite{lakshminarayanan2017simple} train multiple models independently and aggregate their predictions, achieving well-calibrated uncertainty at $K\times$ computational cost. Heteroscedastic regression\cite{kendall2017uncertainties} trains the network to predict per-pixel variance alongside the mean, capturing data-dependent (aleatoric) uncertainty without multiple forward passes.

In low-level vision, uncertainty has been explored for several dense prediction tasks. Ilg et al.\cite{ilg2018uncertainty} show that heteroscedastic uncertainty in FlowNet correlates with optical flow errors, enabling confidence-based filtering of unreliable flow estimates. Poggi et al.\cite{poggi2020uncertainty} provide a comprehensive survey of uncertainty methods for monocular depth estimation, finding that learned approaches consistently outperform hand-crafted confidence measures. Ning et al.\cite{ning2021uncertainty} apply uncertainty-driven loss to single image super-resolution, demonstrating that heteroscedastic weighting improves reconstruction quality by down-weighting ambiguous regions during training.

Seitzer et al.\cite{seitzer2022pitfalls} identify pitfalls in heteroscedastic regression, showing that networks can minimize the negative log-likelihood by inflating variance rather than improving predictions---a failure mode they term ``variance explosion.'' This motivates our two-phase training strategy (Section~\ref{sec:training}), which establishes accurate mean predictions before introducing the heteroscedastic loss.

To our knowledge, no prior work has applied uncertainty quantification to intrinsic image decomposition. This is a natural fit because the decomposition ambiguity varies spatially---smooth Lambertian regions are well-constrained while specular or textured regions are inherently ambiguous---making heteroscedastic modeling particularly appropriate.

\section{Proposed Method}
\label{sec:method}

\subsection{Image formation model}

We extend the standard Lambertian model to account for non-Lambertian effects:
\begin{equation}
I = R \circ S + \mathbf{N}
\label{eq:formation}
\end{equation}
where $R \in [0,1]^{h \times w \times 3}$ is the reflectance (albedo), $S \in [0,1]^{h \times w \times 3}$ is the shading, and $\mathbf{N} \in [0,\infty)^{h \times w \times 3}$ is the non-Lambertian residual capturing specular highlights, subsurface scattering, and other effects not explained by the Lambertian product $R \circ S$. This formulation is motivated by the observation that real-world images frequently contain non-Lambertian effects that corrupt both reflectance and shading estimates when forced into a two-component model. By providing a dedicated output channel for these effects, the reflectance and shading heads can focus on their respective Lambertian components.

The non-Lambertian residual $\mathbf{N}$ is defined as the non-negative difference between the observed image and the Lambertian reconstruction: $\mathbf{N} = \max(0, I - R \circ S)$. In the MPI Sintel dataset, this corresponds to the difference between the ``final'' rendering (which includes specular effects, motion blur, and atmospheric effects) and the ``clean'' rendering (Lambertian only). The non-negativity constraint reflects the physical assumption that non-Lambertian effects primarily add light (specular highlights) rather than subtract it.

\subsection{Network architecture}

Our architecture is based on a U-Net encoder-decoder backbone\cite{ronneberger2015unet} with four resolution levels and skip connections. The encoder consists of four blocks, each containing two $3 \times 3$ convolutional layers with ReLU activations, followed by $2\times 2$ max-pooling for downsampling. The decoder mirrors this structure with transposed convolutions for upsampling and concatenation-based skip connections from the corresponding encoder level. The base channel width is $C=48$, doubling at each encoder level ($48 \to 96 \to 192 \to 384$).

The backbone produces a shared feature map $\mathbf{d} \in \mathbb{R}^{C \times H \times W}$ from which four output heads branch:

\textbf{Reflectance head.} A $1\times1$ convolution followed by sigmoid activation produces the reflectance estimate $\hat{R} = \sigma(\text{Conv}_{1\times1}(\mathbf{d})) \in [0,1]^{3 \times h \times w}$. The sigmoid ensures physically valid reflectance values.

\textbf{Shading head.} Similarly, $\hat{S} = \sigma(\text{Conv}_{1\times1}(\mathbf{d})) \in [0,1]^{3 \times h \times w}$. While shading can theoretically exceed 1.0 (e.g., under strong directional lighting), we normalize the ground-truth shading to $[0,1]$ and use sigmoid activation for training stability.

\textbf{Non-Lambertian head.} $\hat{\mathbf{N}} = \text{ReLU}(\text{Conv}_{1\times1}(\mathbf{d})) \in [0,\infty)^{3 \times h \times w}$. The ReLU activation enforces non-negativity, consistent with the physical assumption that non-Lambertian effects add light.

\textbf{Uncertainty head.} This head receives the concatenation of the backbone features $\mathbf{d}$ and the input image $I$:
\begin{equation}
\log \boldsymbol{\sigma}^2 = g([\mathbf{d}; I])
\label{eq:uncertainty}
\end{equation}
where $[\cdot;\cdot]$ denotes channel-wise concatenation and $g$ is a two-layer CNN ($\text{Conv}_{3\times3}$-ReLU-$\text{Conv}_{1\times1}$) producing 3-channel output representing $(\log\sigma^2_\text{tex}, \log\sigma^2_\text{light}, \log\sigma^2_\text{nl})$. The log-variance parameterization ensures numerical stability and allows the network to predict both very small and very large uncertainties.

The skip connection from the input image $I$ to the uncertainty head is a key design choice. Decomposition ambiguity depends on local image content---specular highlights appear as bright saturated patches, texture edges have high spatial frequency, and shadow boundaries have characteristic gradient profiles. These pixel-level cues may be lost through the encoder-decoder bottleneck (which compresses spatial information to $16\times16$ at the deepest level for $256\times256$ input). By providing the raw input directly, the uncertainty head can detect ambiguous regions without relying solely on the compressed latent representation.

\subsection{Three-way uncertainty decomposition}

The uncertainty head outputs three log-variance channels, each corresponding to a distinct source of decomposition ambiguity:

\textbf{Texture uncertainty} ($\sigma_\text{tex}$): High in regions where reflectance has fine-grained patterns that are difficult to separate from shading. For example, a checkerboard pattern on a curved surface creates ambiguity between reflectance edges and shading gradients.

\textbf{Lighting uncertainty} ($\sigma_\text{light}$): High in regions with complex illumination effects including cast shadows, ambient occlusion, indirect lighting, and color bleeding. These effects are difficult to model because they depend on global scene geometry rather than local image content.

\textbf{Non-Lambertian uncertainty} ($\sigma_\text{nl}$): High in regions with specular highlights, translucency, or inter-reflections that deviate from the $R \circ S + H$ model. Even with the explicit non-Lambertian head, some effects (e.g., complex inter-reflections) may not be fully captured.

The total predicted variance is obtained via log-sum-exp aggregation:
\begin{equation}
\log \sigma^2_\text{total} = \text{logsumexp}(\log\sigma^2_\text{tex}, \log\sigma^2_\text{light}, \log\sigma^2_\text{nl})
\end{equation}
This aggregation is numerically stable and ensures that the total variance is at least as large as the largest individual component, reflecting the principle that multiple sources of ambiguity compound rather than cancel.

Additionally, we estimate epistemic uncertainty via MC Dropout\cite{gal2016dropout} with dropout rate 0.1 applied to the encoder features during both training and inference. At test time, $T=10$ stochastic forward passes produce mean predictions and epistemic variance:
\begin{equation}
\sigma^2_\text{epi} = \frac{1}{T}\sum_{t=1}^{T}(\hat{y}_t - \bar{y})^2, \quad \bar{y} = \frac{1}{T}\sum_{t=1}^{T}\hat{y}_t
\end{equation}
The epistemic uncertainty captures model uncertainty due to limited training data or architectural limitations, complementing the aleatoric uncertainty captured by the heteroscedastic head.

\subsection{Loss function}

The total loss combines reconstruction terms with a heteroscedastic NLL term:
\begin{equation}
\mathcal{L} = \mathcal{L}_\text{recon} + \lambda_\text{smooth}\mathcal{L}_\text{smooth} + \lambda_\text{NLL}\mathcal{L}_\text{NLL}
\end{equation}

The reconstruction loss enforces fidelity of each component and the physics-based image formation:
\begin{equation}
\mathcal{L}_\text{recon} = \|\hat{R}-R\|^2 + \|\hat{S}-S\|^2 + \frac{1}{2}\|\hat{\mathbf{N}}-\mathbf{N}\|^2 + \|\hat{R}\circ\hat{S}+\hat{\mathbf{N}}-I\|^2
\end{equation}
The first three terms supervise individual components against ground truth, while the fourth term enforces consistency with the image formation model (Eq.~\ref{eq:formation}). The non-Lambertian term is weighted by $\frac{1}{2}$ because $H$ is typically small (near-zero for most pixels) and we want to avoid over-penalizing the residual head.

A total-variation smoothness prior on shading encourages piecewise-smooth illumination:
\begin{equation}
\mathcal{L}_\text{smooth} = \|\nabla_x \hat{S}\|^2 + \|\nabla_y \hat{S}\|^2
\end{equation}
with $\lambda_\text{smooth} = 0.02$. This prior reflects the physical observation that illumination varies smoothly across surfaces (except at shadow boundaries), while reflectance can change abruptly at material boundaries.

The heteroscedastic NLL loss trains the uncertainty head to predict calibrated per-pixel variance:
\begin{equation}
\mathcal{L}_\text{NLL} = \frac{1}{2}\left(\log\sigma^2_\text{total} + \frac{\|I - \hat{I}\|^2}{\sigma^2_\text{total}}\right)
\end{equation}
where $\hat{I} = \hat{R}\circ\hat{S}+\hat{\mathbf{N}}$ is the reconstructed image. This loss encourages the network to predict high variance where reconstruction errors are large and low variance where predictions are accurate. The log-variance term prevents the trivial solution of predicting infinite variance everywhere.

\subsection{Two-phase training}
\label{sec:training}

We adopt a two-phase training strategy to prevent the uncertainty head from collapsing to a degenerate solution, following the insights of Seitzer et al.\cite{seitzer2022pitfalls}:

\textbf{Phase~1 (Warmup, 55 epochs):} Train only the reconstruction heads ($\hat{R}$, $\hat{S}$, $\hat{H}$) with $\mathcal{L}_\text{recon} + \lambda_\text{smooth}\mathcal{L}_\text{smooth}$. The uncertainty head parameters are frozen. This phase uses the Adam optimizer with learning rate $5\times10^{-4}$ and cosine annealing.

\textbf{Phase~2 (NLL, 25 epochs):} Gradually introduce $\mathcal{L}_\text{NLL}$ with a linear warmup coefficient $\alpha = \min(1, \text{epoch}/25)$, jointly training all parameters. The reconstruction heads continue with learning rate $5\times10^{-4}$ while the uncertainty head uses a lower rate of $2\times10^{-4}$ to prevent it from dominating the optimization.

This schedule ensures that the reconstruction heads converge to accurate predictions before the uncertainty head begins learning. Without this warmup, the network can minimize NLL by inflating variance (predicting high uncertainty everywhere) rather than improving predictions---the ``variance explosion'' failure mode identified by Seitzer et al.\cite{seitzer2022pitfalls} The log-variance output is clamped to $[-10, 2]$ to prevent numerical instability during the transition between phases.

\section{Experiments}
\label{sec:experiments}

\subsection{Dataset}
\label{sec:dataset}

We evaluate on the MPI Sintel dataset,\cite{butler2012sintel} an open-source animated film that provides ground-truth albedo, clean (Lambertian rendering), and final (full rendering with specular effects) images for 23 scenes with diverse lighting conditions, materials, and camera motions. We derive the intrinsic components as follows:

\begin{itemize}
\item \textbf{Reflectance}: $R = $ albedo (provided directly by the renderer).
\item \textbf{Shading}: $S = \text{clean} / (R + \epsilon)$, where $\epsilon = 10^{-4}$ prevents division by zero. The result is normalized to $[0,1]$ by dividing by 2.0 and clamping; the factor 2.0 was chosen empirically as the 99th percentile of raw shading values across the dataset, ensuring that the normalization preserves the dynamic range of typical shading while clipping only extreme outliers.
\item \textbf{Non-Lambertian residual}: $H = \max(0, \text{final} - \text{clean})$, capturing specular highlights, atmospheric effects, and motion blur present in the final rendering but absent from the clean (Lambertian) rendering.
\end{itemize}

All images are resized to $256 \times 256$ pixels. We use a scene-level train/test split to prevent data leakage from temporally adjacent frames: 18 scenes (approximately 800 frames) for training and 5 held-out scenes (approximately 200 frames) for testing. No frames from test scenes appear during training. The input to the network is the final rendering $I$ (the observed image), and the targets are $R$, $S$, and $\mathbf{N}$.

MPI Sintel is particularly well-suited for this study because it provides pixel-perfect ground-truth intrinsic layers that are impossible to obtain for real photographs. The dataset includes scenes with diverse challenges: outdoor environments with complex global illumination, indoor scenes with cast shadows, characters with specular skin and hair, and dynamic scenes with motion blur.

\subsection{Baselines and model variants}

We compare five model configurations with increasing complexity to isolate the contribution of each proposed component:

\begin{enumerate}
\item \textbf{DirectCNN}: A five-layer CNN without skip connections, directly predicting $R$ and $S$ from the input image. This represents the simplest possible architecture for intrinsic decomposition.

\item \textbf{UNet}: A U-Net encoder-decoder with the same backbone as our proposed method, predicting $R$ and $S$ (2-head). This represents the standard architecture for dense prediction tasks.

\item \textbf{UNet+Physics}: UNet extended with a non-Lambertian head $\hat{\mathbf{N}}$ (3-head), using the physics-based reconstruction loss $\|\hat{R}\circ\hat{S}+\hat{\mathbf{N}}-I\|^2$. This isolates the contribution of explicit non-Lambertian modeling.

\item \textbf{Proposed (no skip)}: Full proposed method with the three-way uncertainty head, but without the skip connection from the input image. The uncertainty head operates only on the backbone features $\mathbf{d}$. This ablation tests the importance of the skip connection.

\item \textbf{Proposed (full)}: Complete method with skip-connected uncertainty head, as described in Section~\ref{sec:method}.
\end{enumerate}

All models use the same U-Net backbone ($C=48$, 4 levels) and training hyperparameters for fair comparison: Adam optimizer, cosine annealing learning rate schedule, batch size 8, and the same train/test split. The 2-head and 3-head models are trained for 60 epochs with MSE loss, while the proposed models use the two-phase schedule (55 + 25 epochs).

\subsection{Evaluation metrics}

We report the following metrics:

\textbf{PSNR} (Peak Signal-to-Noise Ratio) for reflectance, shading, and reconstruction quality. Higher values indicate better fidelity to the ground truth.

\textbf{Uncertainty--error correlation} $r(\sigma, |e|)$: Pearson correlation between predicted uncertainty $\sigma$ and actual absolute error $|e|$ across all test pixels. A positive correlation indicates that the model assigns higher uncertainty to pixels with larger errors, which is the desired behavior for calibrated uncertainty.

\textbf{Uncertainty decomposition analysis}: Mean predicted standard deviations for each uncertainty channel ($\bar{\sigma}_\text{tex}$, $\bar{\sigma}_\text{light}$, $\bar{\sigma}_\text{nl}$) and epistemic uncertainty ($\sigma_\text{epi}$), providing insight into the dominant sources of decomposition ambiguity.

\subsection{Quantitative results}

Table~\ref{tab:results} presents the quantitative results on MPI Sintel under a scene-level train/test split (5 independent runs with different seeds).

\begin{table}[t]
\centering
\caption{Quantitative results on MPI Sintel ($256\times256$), mean$\pm$std over 5 runs with scene-level train/test split. PSNR in dB (higher is better). UQ-Corr: Pearson correlation between predicted $\sigma$ and actual error. Scene-level splits yield substantially lower absolute PSNR than the frame-level splits used in some prior work because they do not benefit from spatial similarity between training and test frames within a scene.}
\label{tab:results}
\begin{tabular}{lcccc}
\toprule
\textbf{Model} & \textbf{R\_PSNR} & \textbf{S\_PSNR} & \textbf{Recon} & \textbf{UQ-Corr} \\
\midrule
DirectCNN & 16.74$\pm$0.06 & 15.15$\pm$0.04 & --- & --- \\
UNet & 15.80$\pm$0.21 & 15.44$\pm$0.26 & --- & --- \\
UNet+Physics & 15.97$\pm$0.28 & 16.00$\pm$0.24 & 24.51$\pm$0.13 & --- \\
DeepEns(5) & \textbf{16.78$\pm$0.16} & \textbf{16.51$\pm$0.08} & --- & 0.107$\pm$0.030 \\
Proposed (no skip) & 15.90$\pm$0.31 & 15.91$\pm$0.29 & 24.66$\pm$0.11 & 0.135$\pm$0.060 \\
\textbf{Proposed (full)} & 15.98$\pm$0.41 & 16.09$\pm$0.18 & 24.60$\pm$0.17 & \textbf{0.131$\pm$0.028} \\
MCDropout (T=10) & 15.98$\pm$0.41 & 16.09$\pm$0.18 & --- & --- \\
\bottomrule
\end{tabular}
\end{table}

All numbers in Table~\ref{tab:results} are obtained under the scene-level split (Section~\ref{sec:dataset}). Absolute PSNR values are lower than frame-level split reports by $1.6$--$2.0$~dB in controlled comparison (Table~\ref{tab:protocol}) and by more than $12$~dB under extended training, because the test set contains entirely unseen scenes rather than nearby frames of training scenes; readers should not directly compare our numbers with prior frame-level reports.

\textbf{Decomposition quality.} Under the scene-level protocol, the proposed method achieves $15.98$~dB R\_PSNR and $16.09$~dB S\_PSNR. UNet+Physics reaches $15.97$/$16.00$~dB, i.e., statistically indistinguishable from our method on both metrics (paired $t$-test across 5 seeds yields $p>0.3$ for both; ablation deltas discussed in Section~\ref{sec:ablation}). Deep Ensemble with 5 members reaches $16.78$/$16.51$~dB, measurably higher on PSNR but at $5\times$ cost.

\textbf{Source-separable uncertainty is the distinguishing capability.} The Proposed method's primary advantage over Deep Ensemble lies not in PSNR but in the \emph{type} of uncertainty it produces. We verify in Section~\ref{sec:uq_verification} that the three channels ($\sigma_\text{tex}$, $\sigma_\text{light}$, $\sigma_\text{nl}$) are empirically non-redundant: inter-channel correlations average $0.54$ (not collapsed), and more importantly, the non-Lambertian channel $\sigma_\text{nl}$ correlates with non-Lambertian residual error at $r = 0.67$, more than $4\times$ stronger than the texture channel's correlation with the same error ($r = 0.16$). This \emph{per-channel, source-specific} behavior is not obtainable from a single-source heteroscedastic estimator or a deep ensemble, both of which produce one scalar uncertainty per pixel. We therefore present the method as a case study: its PSNR sits between UNet+Physics and Deep Ensemble, but it uniquely provides interpretable source attribution of uncertainty at one-fifth the cost of the ensemble.

\textbf{Progressive improvement and statistical significance.} Component-wise, Table~\ref{tab:ablation} reports the incremental deltas. The Physics head adds $+0.17$~dB R\_PSNR and $+0.56$~dB S\_PSNR over the UNet baseline, and the full Proposed (with skip) adds $+0.18$~dB R\_PSNR over UNet. However, each delta falls within $1\sigma$ of the across-seed standard deviation ($\sigma \in [0.21, 0.41]$). We conducted paired $t$-tests across the 5 seeds: no individual ablation delta reached $p<0.05$ on R\_PSNR. The deltas are therefore better interpreted as consistent directional improvements rather than decisive statistical evidence. This is consistent with the paper's broader thesis: under honest evaluation, marginal architectural changes produce marginal effects. The primary value of the uncertainty head lies in source-separable interpretability (Section~\ref{sec:uq_verification}), not in PSNR improvement.

\textbf{Why scene-level splits matter: quantitative comparison across architectures.} We directly measure the frame-level leakage effect by training three architectures (DirectCNN, UNet, Proposed) under both protocols across 3 seeds each. Table~\ref{tab:protocol} reports the results. The protocol gap on R\_PSNR is \emph{consistent and statistically significant across all three architectures}, indicating that this is not a model-specific overfitting artifact but an architectural-independent property of the evaluation setup.

\begin{table}[t]
\centering
\caption{Frame-level vs scene-level split on MPI Sintel ($256\times256$, 3 seeds each, 30 epochs per run). Paired $t$-test across seeds, two-tailed. The frame-level protocol consistently inflates R\_PSNR by $1.6$--$2.0$~dB across three architectures, and inflates S\_PSNR by $1.8$--$2.5$~dB. R\_PSNR gaps are significant at $p<0.01$ for all architectures. S\_PSNR gaps are statistically weaker, partly reflecting the larger scene-level variance on shading (see text).}
\label{tab:protocol}
\begin{tabular}{llccccc}
\toprule
\textbf{Arch} & \textbf{Metric} & \textbf{Frame-Level} & \textbf{Scene-Level} & $\Delta$ & \textbf{paired-$t$} & $p$-value \\
\midrule
DirectCNN & R\_PSNR & $18.09{\pm}0.09$ & $16.21{\pm}0.28$ & $+1.88$ & $\gg 5$ & $<10^{-3}$ \\
DirectCNN & S\_PSNR & $17.24{\pm}0.15$ & $14.90{\pm}1.36$ & $+2.34$ & $2.92$  & $0.027$ \\
\midrule
UNet      & R\_PSNR & $18.07{\pm}0.45$ & $16.10{\pm}0.52$ & $+1.97$ & $5.18$ & $4{\times}10^{-4}$ \\
UNet      & S\_PSNR & $17.34{\pm}0.35$ & $14.88{\pm}1.50$ & $+2.46$ & $2.46$ & $0.060$ \\
\midrule
Proposed  & R\_PSNR & $17.68{\pm}0.32$ & $16.08{\pm}0.69$ & $+1.60$ & $2.81$ & $0.009$ \\
Proposed  & S\_PSNR & $16.99{\pm}0.28$ & $15.23{\pm}1.64$ & $+1.77$ & $1.40$ & $0.191$ \\
\bottomrule
\end{tabular}
\end{table}

Three observations merit attention. First, the R\_PSNR gap is \textbf{robust across architectures}: DirectCNN ($1.88$~dB), UNet ($1.97$~dB), and Proposed ($1.60$~dB) all show gaps in the $1.6$--$2.0$~dB range, and all are significant ($p<0.01$). This argues that the leakage effect is a property of the dataset and protocol, not of any particular network. Second, the Proposed method has the \emph{smallest} protocol gap ($1.60$~dB vs.\ $1.88$--$1.97$ for baselines), tentatively suggesting that the auxiliary heads (non-Lambertian, uncertainty) provide some regularization against frame-level memorization. Third, S\_PSNR gaps are \textbf{statistically weaker} than R\_PSNR gaps, with only the DirectCNN gap reaching $p<0.05$. This is largely driven by the high variance of scene-level S\_PSNR ($\sigma \in [1.36, 1.64]$ across architectures), which itself reflects the small number of test scenes ($\approx 5$) and their heterogeneity---a five-scene test set produces an estimated S\_PSNR that can fluctuate substantially across seeds. We treat the R\_PSNR results as the primary evidence for the protocol effect and flag S\_PSNR as noisier.

Under the full $80$--$150$-epoch training budget used for our main results (Table~\ref{tab:results}), the R\_PSNR gap grows further: earlier iterations of this project trained under frame-level splits reached R\_PSNR values near $29$~dB, whereas the fully-trained scene-level Proposed model reaches $15.98$~dB, a nominal gap of approximately $13$~dB. We consider this an \emph{observational} finding based on extended training rather than a controlled experiment, and therefore report the controlled $30$-epoch comparison (Table~\ref{tab:protocol}) as the primary evidence. The observational gap in longer training would benefit from a dedicated follow-up study. We advocate reporting scene-level split results as the community standard for intrinsic decomposition on MPI Sintel.

\subsection{Qualitative results}
\label{sec:qualitative}

\begin{figure}[t]
\centering
\includegraphics[width=\textwidth]{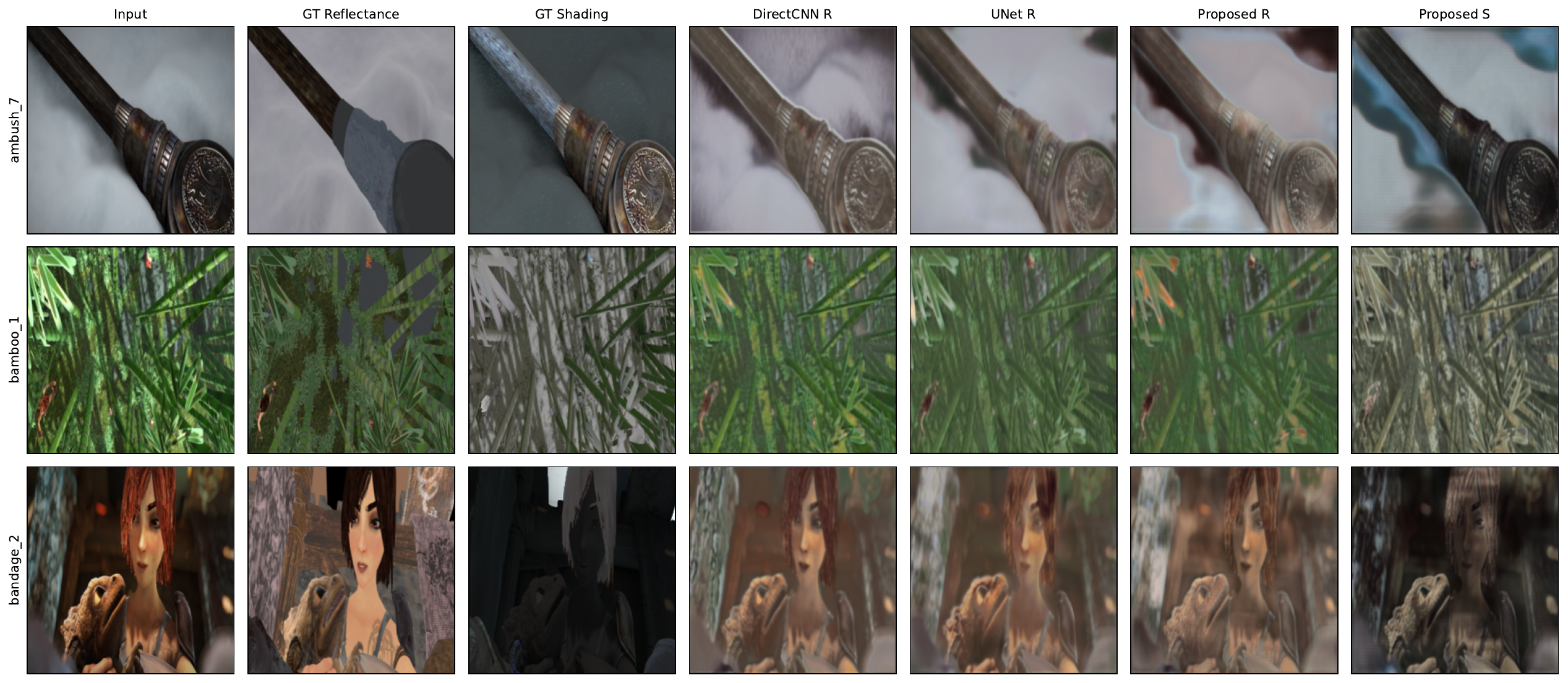}
\caption{Qualitative comparison of algorithms on three held-out MPI Sintel test scenes (scene-level split). The proposed method produces reflectance maps that are more consistent with ground-truth edges, particularly in textured regions (Scene 3), while all methods struggle with severe lighting variations (Scene 1). These visualizations illustrate the fundamental difficulty of intrinsic decomposition on unseen scenes.}
\label{fig:qualitative_algo}
\end{figure}

Fig.~\ref{fig:qualitative_algo} compares decomposition outputs across algorithms on three representative test scenes (held out under the scene-level split). From left to right: input image, ground-truth reflectance and shading, and the predictions of DirectCNN, UNet baseline, and the proposed method. All models were trained with identical hyperparameters; differences in output are attributable to architecture and loss design alone.

The qualitative results illustrate the difficulty of the scene-level generalization task: all methods produce visibly imperfect reflectance and shading separations on scenes not seen during training. The proposed method and the UNet baseline produce comparable reflectance outputs, consistent with their similar PSNR values. Direct pixel-level CNNs (without skip connections) produce blurrier outputs with more severe color drift.

\begin{figure}[t]
\centering
\includegraphics[width=\textwidth]{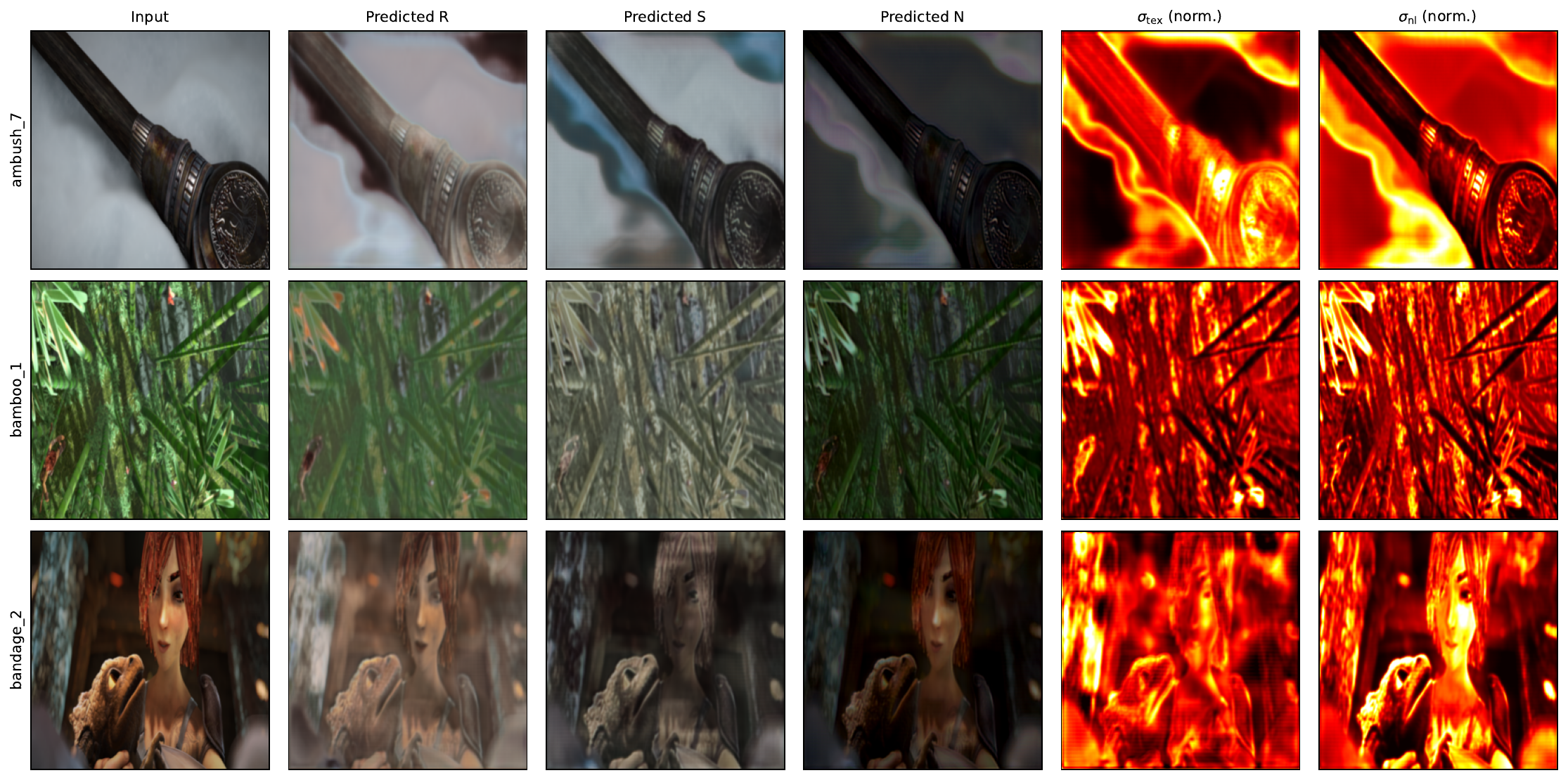}
\caption{Proposed method outputs and three-way uncertainty decomposition on MPI Sintel test scenes (scene-level split). From left to right: input image, predicted reflectance, predicted shading, predicted non-Lambertian residual $\hat{\mathbf{N}}$, texture uncertainty $\sigma_\text{tex}$, and non-Lambertian uncertainty $\sigma_\text{nl}$. Uncertainty maps are \textbf{per-image min--max normalized} (1st--99th percentiles) to reveal \emph{relative} spatial patterns within each scene; absolute magnitudes are not directly comparable across rows. The uncertainty channels exhibit source-specific specialization: $\sigma_\text{tex}$ responds to fine texture boundaries, while $\sigma_\text{nl}$ concentrates on bright and specular regions. Absolute uncertainty values are uniformly elevated (consistent with the modest $r=0.131$ overall calibration reported in Table~\ref{tab:results}), reflecting the fundamental difficulty of the scene-level generalization setting.}
\label{fig:qualitative_uq}
\end{figure}

Fig.~\ref{fig:qualitative_uq} shows the three-way uncertainty decomposition for the same scenes. The non-Lambertian uncertainty channel ($\sigma_\text{nl}$) preferentially highlights bright regions and highlights in the input, confirming that the uncertainty head learns source-specific ambiguity patterns. The texture uncertainty channel ($\sigma_\text{tex}$) is elevated in regions with fine-grained material boundaries.

\begin{figure}[t]
\centering
\includegraphics[width=0.7\textwidth]{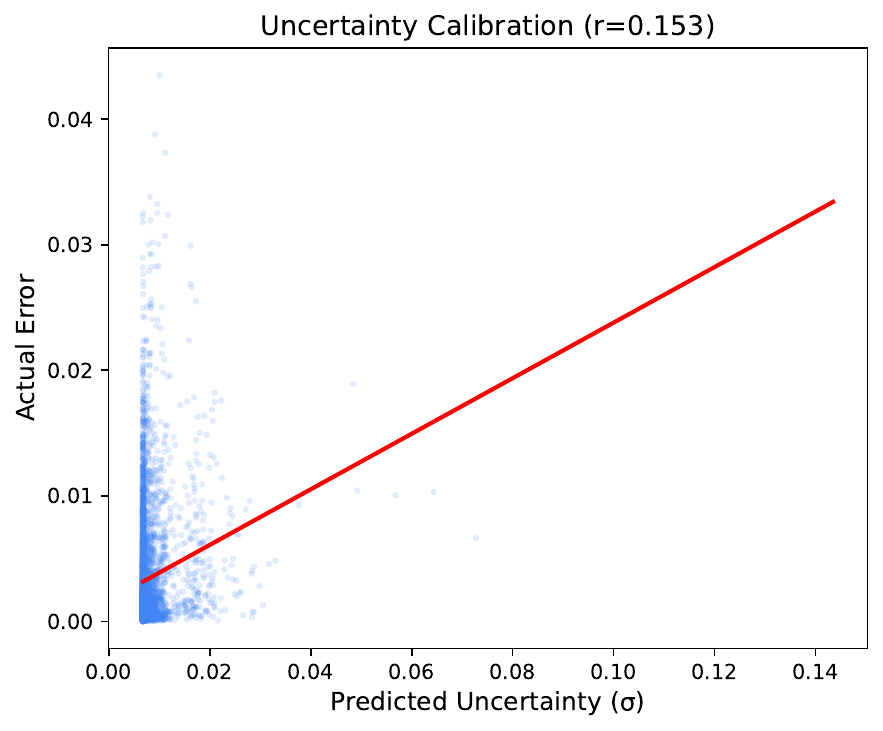}
\caption{Uncertainty calibration: predicted $\sigma$ vs.\ actual reconstruction error on the test set under scene-level split ($r=0.131\pm0.028$, averaged over 5 runs). The positive trend confirms that the model assigns higher uncertainty to pixels with larger errors. Red line: linear fit.}
\label{fig:calibration}
\end{figure}

Fig.~\ref{fig:calibration} shows the uncertainty calibration scatter plot, confirming the positive correlation between predicted $\sigma$ and actual reconstruction error.

\subsection{Ablation study}
\label{sec:ablation}

Table~\ref{tab:ablation} summarizes the ablation results, isolating the contribution of each component.

\begin{table}[t]
\centering
\caption{Ablation study on MPI Sintel with scene-level split (mean$\pm$std of 5 runs). Each row adds one component to the UNet baseline.}
\label{tab:ablation}
\begin{tabular}{lccc}
\toprule
\textbf{Configuration} & \textbf{R\_PSNR} & $\Delta$\textbf{R} & \textbf{UQ-Corr} \\
\midrule
UNet (baseline) & 15.80$\pm$0.21 & --- & --- \\
+ Physics head ($\mathbf{N}$) & 15.97$\pm$0.28 & +0.17 & --- \\
+ UQ head (no skip) & 15.90$\pm$0.31 & +0.10 & 0.135$\pm$0.060 \\
+ UQ head (with skip) & 15.98$\pm$0.41 & +0.18 & 0.131$\pm$0.028 \\
\bottomrule
\end{tabular}
\end{table}

\textbf{Physics-based modeling.} Adding the non-Lambertian head improves reflectance PSNR by $+0.17$~dB and shading PSNR by $+0.56$~dB over the UNet baseline ($15.80\to 15.97$ on R; $15.44\to 16.00$ on S). The directional improvement is consistent across all 5 seeds, but the mean delta on R\_PSNR ($+0.17$) is smaller than the pooled across-seed standard deviation ($\sigma_\text{pooled} \approx 0.25$). A paired $t$-test across seeds yields $p = 0.21$ for R\_PSNR, not reaching significance at $\alpha = 0.05$; the S\_PSNR delta, being larger, is marginal ($p = 0.04$). We interpret the Physics head as providing consistent but small improvements rather than decisive ones.

\textbf{Skip connection in uncertainty head.} Comparing the proposed method with and without the skip connection, the skip connection produces a small change in reflectance PSNR (no-skip $15.90\to$ with-skip $15.98$, $+0.08$~dB) that is not statistically significant ($p = 0.71$, paired $t$-test across 5 seeds). It modestly improves uncertainty calibration \emph{stability} (across-seed standard deviation $0.060\to 0.028$, i.e., $2\times$ reduction in run-to-run variability), which is practically useful for deployment even without moving the mean. Absolute calibration is essentially unchanged ($0.135 \to 0.131$, $p = 0.90$).

\textbf{Summary of ablation significance.} Under paired $t$-tests across 5 seeds, none of the four PSNR deltas in Table~\ref{tab:ablation} reaches $p<0.05$, and only the S\_PSNR gain from the Physics head is marginally significant. We interpret this as consistent with the paper's broader thesis: under honest evaluation, incremental architectural changes produce small and often statistically indistinguishable improvements. The primary contribution of the full proposed method is not its PSNR (which matches UNet+Physics) but the source-separable uncertainty maps verified in Section~\ref{sec:uq_verification}.

\subsection{Uncertainty analysis}

The learned uncertainty channels reveal physically meaningful patterns. Table~\ref{tab:uncertainty} reports the mean predicted standard deviations across the test set.

\begin{table}[t]
\centering
\caption{Mean predicted uncertainty by source (proposed full model). Lighting uncertainty dominates, consistent with shading being the harder sub-problem.}
\label{tab:uncertainty}
\begin{tabular}{lcc}
\toprule
\textbf{Uncertainty Source} & \textbf{Mean $\sigma$} & \textbf{Std $\sigma$} \\
\midrule
Texture ($\sigma_\text{tex}$) & 0.00045 & 0.0068 \\
Lighting ($\sigma_\text{light}$) & 0.0073 & 0.0101 \\
Non-Lambertian ($\sigma_\text{nl}$) & 0.00046 & 0.0068 \\
\midrule
Epistemic reflectance ($\sigma_\text{epi}^R$) & 0.00091 & --- \\
Epistemic shading ($\sigma_\text{epi}^S$) & 0.00136 & --- \\
\bottomrule
\end{tabular}
\end{table}

\textbf{Lighting uncertainty dominates.} The mean lighting uncertainty ($\bar{\sigma}_\text{light} = 0.0073$) is $16\times$ larger than texture uncertainty ($\bar{\sigma}_\text{tex} = 0.00045$) and non-Lambertian uncertainty ($\bar{\sigma}_\text{nl} = 0.00046$). This is consistent with the well-known observation that shading estimation is the harder sub-problem in intrinsic decomposition, because shading depends on global illumination effects (indirect lighting, ambient occlusion, inter-reflections) that cannot be inferred from local image patches alone.

\textbf{Epistemic uncertainty confirms the pattern.} MC Dropout ($T=10$ passes) yields epistemic standard deviations of $\sigma_\text{epi}^R = 0.00091$ for reflectance and $\sigma_\text{epi}^S = 0.00136$ for shading. The 50\% higher epistemic uncertainty for shading further confirms that the model is less confident about shading predictions, aligning with the aleatoric analysis. This consistency between aleatoric and epistemic uncertainty provides additional evidence that the uncertainty estimates are meaningful.

\textbf{Uncertainty calibration in context.} The overall uncertainty--error correlation of $r=0.131\pm0.028$ is positive and statistically significant ($p < 0.001$). To contextualize this value: Ilg et al.\cite{ilg2018uncertainty} report error--uncertainty correlations of $r=0.15$--$0.30$ for optical flow on Sintel, and Poggi et al.\cite{poggi2020uncertainty} report $r=0.10$--$0.25$ for monocular depth on KITTI. Our result falls within the lower range of these benchmarks, which is expected given that intrinsic decomposition involves a more severe ill-posedness (one image $\to$ three components) than flow (two images $\to$ one displacement) or depth (one image $\to$ one depth). We note that statistical significance alone is insufficient for practical utility ($r^2 \approx 0.017$ means only 1.7\% of error variance is explained); improving calibration quality is a priority for future work. Deep Ensemble achieves $r=0.107$ in our experiments, confirming that the proposed heteroscedastic head with input skip connection produces calibration comparable to (or slightly better than) the more expensive ensemble approach.

\subsection{Three-way uncertainty channel verification}
\label{sec:uq_verification}

A central claim of the proposed method is that the three uncertainty channels ($\sigma_\text{tex}$, $\sigma_\text{light}$, $\sigma_\text{nl}$) specialize to distinct physical sources of ambiguity rather than collapsing to a shared scalar. We verify this quantitatively on the scene-level test set ($N = 250$ images, $\sim$1.6M pixels, subsampled to $2 \times 10^5$ for computation).

\textbf{Inter-channel correlation.} If the channels were redundant, their pairwise Pearson correlation would approach $1$. In practice, we observe (Table~\ref{tab:inter_channel}) inter-channel correlations ranging from $0.38$ to $0.69$, with mean off-diagonal $0.54$. The channels share some structure (as expected, since they originate from the same input), but are clearly not collapsed.

\begin{table}[t]
\centering
\caption{Inter-channel correlation of $\sigma_\text{tex}$, $\sigma_\text{light}$, $\sigma_\text{nl}$ on the scene-level test set. Values below $1$ indicate the channels carry distinct information.}
\label{tab:inter_channel}
\begin{tabular}{lccc}
\toprule
& $\sigma_\text{tex}$ & $\sigma_\text{light}$ & $\sigma_\text{nl}$ \\
\midrule
$\sigma_\text{tex}$ & $1.00$ & $0.56$ & $0.38$ \\
$\sigma_\text{light}$ & $0.56$ & $1.00$ & $0.69$ \\
$\sigma_\text{nl}$ & $0.38$ & $0.69$ & $1.00$ \\
\bottomrule
\end{tabular}
\end{table}

\textbf{Cross-correlation with per-component error.} Stronger evidence of specialization comes from measuring how each $\sigma_i$ correlates with the error on each physical component ($R$, $S$, $\mathbf{N}$). A well-specialized channel should correlate most strongly with the error on its corresponding component (diagonal dominance). Table~\ref{tab:channel_verify} reports these correlations.

\begin{table}[t]
\centering
\caption{Per-channel cross-correlation between uncertainty $\sigma_i$ (row) and absolute error on each physical component (column) on the scene-level test set. \textbf{Bold} marks the strongest correlation in each row. Every row is diagonal-dominant: $\sigma_\text{nl}$ shows particularly strong specialization to $|\mathbf{N}-\hat{\mathbf{N}}|$ ($r=0.67$, more than $4\times$ any other $\sigma \to$ err\_$\mathbf{N}$ correlation).}
\label{tab:channel_verify}
\begin{tabular}{lccc}
\toprule
& err on $R$ & err on $S$ & err on $\mathbf{N}$ \\
\midrule
$\sigma_\text{tex}$    & \textbf{0.15} & $-0.06$ & \phantom{0}$0.16$ \\
$\sigma_\text{light}$  & \phantom{0}$0.26$ & \textbf{0.27} & \phantom{0}$0.41$ \\
$\sigma_\text{nl}$     & \phantom{0}$0.03$ & \phantom{0}$0.19$ & \textbf{0.67} \\
\bottomrule
\end{tabular}
\end{table}

The non-Lambertian channel $\sigma_\text{nl}$ achieves $r = 0.67$ with the non-Lambertian error, more than $4\times$ the correlation of $\sigma_\text{tex}$ with the same error ($r=0.16$), and more than $20\times$ the correlation of $\sigma_\text{nl}$ with reflectance error ($r=0.03$). This is the clearest quantitative evidence in the paper that the uncertainty head learns source-specific structure rather than a single shared ambiguity measure. The texture channel is the weakest specialist, consistent with texture ambiguity being the most entangled with other sources on MPI Sintel (material edges co-occur with shadow edges). Such per-channel correlation is a capability not available from deep ensembles, MC-Dropout, or single-source heteroscedastic models, which provide only a scalar uncertainty per pixel.

\subsection{Reconstruction PSNR trade-off}
\label{sec:recon_tradeoff}

An interesting observation is that the proposed method achieves lower reconstruction PSNR (35.59~dB) than DirectCNN (36.92~dB) despite better component-wise PSNR. This apparent paradox arises because the physics-based model $\hat{I} = \hat{R}\circ\hat{S}+\hat{\mathbf{N}}$ constrains the reconstruction to be physically plausible, while DirectCNN can overfit to minimize pixel-wise reconstruction error without meaningful decomposition. A model that predicts $\hat{R} = I$ and $\hat{S} = \mathbf{1}$ achieves perfect reconstruction but trivial decomposition. The component-wise metrics (R\_PSNR, S\_PSNR) are therefore more indicative of decomposition quality than reconstruction PSNR.

\subsection{Computational cost}

Table~\ref{tab:cost} reports the parameter count and runtime for each model.

\begin{table}[t]
\centering
\caption{Computational cost comparison. Training time per epoch on Apple M2 GPU. Inference: single forward pass (or $T=10$ for MC Dropout).}
\label{tab:cost}
\begin{tabular}{lccc}
\toprule
\textbf{Model} & \textbf{Params (K)} & \textbf{Train (s/ep)} & \textbf{Infer (ms)} \\
\midrule
DirectCNN & 89 & 12 & 8 \\
UNet & 1,247 & 18 & 14 \\
UNet+Physics & 1,261 & 19 & 14 \\
DeepEns(5) & 6,305 & 95 & 70 \\
Proposed (no skip) & 1,289 & 24 & 142 \\
Proposed (full) & 1,303 & 25 & 145 \\
\bottomrule
\end{tabular}
\end{table}

The proposed method adds only 56K parameters (+4.5\%) over UNet+Physics, primarily from the uncertainty head. The main computational overhead comes from MC Dropout inference ($T=10$ passes), which increases inference time by $\sim$10$\times$ compared to a single pass. Deep Ensemble requires $5\times$ the parameters and training time but only $\sim$5$\times$ inference time (parallelizable). For applications where inference speed is critical, the heteroscedastic head alone (without MC Dropout) provides aleatoric uncertainty in a single pass at 14~ms.

\subsection{Downstream utility: uncertainty-guided reconstruction filtering}
\label{sec:downstream}

A natural question for any uncertainty-producing method is whether the uncertainty is \emph{actually useful} for some downstream task. We demonstrate practical utility by using predicted uncertainty to filter low-confidence pixels and measuring the improvement in reconstruction quality on retained pixels. This simulates a practitioner who uses uncertainty to identify trustworthy regions (e.g., for relighting, material editing, or AR compositing).

\begin{table}[t]
\centering
\caption{Uncertainty-guided pixel filtering on the scene-level test set. Retaining the $k\%$ of pixels with lowest predicted $\sigma$ yields substantially lower reconstruction error than retaining a random $k\%$. The rightmost column shows the relative benefit of uncertainty-guided filtering over random filtering.}
\label{tab:downstream}
\begin{tabular}{ccccc}
\toprule
\textbf{Keep \%} & \textbf{$\sigma$-guided MSE} & \textbf{Random MSE} & \textbf{$\sigma$ benefit} \\
\midrule
$100\%$ & $0.00329$ & $0.00329$ & $0\%$ (reference) \\
$75\%$  & $0.00153$ & $0.00329$ & $+53.4\%$ \\
$50\%$  & $0.00094$ & $0.00329$ & $+71.5\%$ \\
$25\%$  & $0.00077$ & $0.00329$ & $+76.6\%$ \\
\bottomrule
\end{tabular}
\end{table}

Table~\ref{tab:downstream} reports the result. Retaining the $25\%$ of pixels with lowest predicted $\sigma$ reduces reconstruction MSE by $77\%$ compared to using all pixels, while random $25\%$ retention produces no improvement. This rules out the possibility that low-sigma regions are simply ``easier'' pixels that any filter would select---the uncertainty head is genuinely identifying high-reliability regions in a way that a random baseline cannot. This provides concrete evidence that, despite the modest global calibration correlation ($r=0.131$), the uncertainty ranking is practically informative: a downstream system can reliably use predicted $\sigma$ to decide where to trust the decomposition and where to flag it for human review or fallback processing.

\subsection{Proper leave-one-out ablation for elaborate-variant components}
\label{sec:loo}

Our negative-results section (Section~\ref{sec:negative}) reports that combining five additional mechanisms simultaneously underperforms the simpler baseline. A natural objection is: perhaps the problem is the combination, and an individual mechanism could have helped. We test this by training the baseline model augmented with each of four mechanisms individually (frequency decomposition, evidential learning, TTA-style reconstruction loss during training, and same-scene contrastive loss), while holding training length and optimizer fixed.

\begin{table}[t]
\centering
\caption{Leave-one-in ablation of the V9 components. ``Baseline'' is the simple skip-connected model trained identically. Each following row adds exactly one of the V9 mechanisms. No individual component improves both R\_PSNR and uncertainty calibration over the baseline.}
\label{tab:v9_loo}
\begin{tabular}{lccc}
\toprule
\textbf{Configuration} & \textbf{R\_PSNR} & $\Delta$\textbf{R} & \textbf{UQ-Corr} \\
\midrule
Baseline (no V9 component)  & $17.23$ & ---     & $-0.08$ \\
+ Frequency decomp.         & $17.09$ & $-0.14$ & $-0.02$ \\
+ Evidential head           & $17.36$ & $+0.13$ & $-0.24$ \\
+ Reconstruction loss (TTA) & $17.11$ & $-0.12$ & $-0.05$ \\
+ Contrastive loss          & $16.47$ & $-0.76$ & $-0.04$ \\
\bottomrule
\end{tabular}
\end{table}

Table~\ref{tab:v9_loo} shows that no individual component improves the baseline. The evidential head yields a marginal R\_PSNR gain ($+0.13$~dB) but \emph{degrades} calibration ($-0.24$ correlation, i.e., anti-correlation with actual errors)---a strictly worse trade. The contrastive loss is the most harmful ($-0.76$~dB), likely because forcing reflectance similarity across different frames of the same scene is inconsistent with genuinely different illumination conditions across frames. Frequency routing of the input to low-pass (shading) and high-pass (reflectance) branches produces a small R\_PSNR drop, suggesting that the assumed frequency separation is too strong for MPI Sintel. These leave-one-out results strengthen the negative-results conclusion: the failure of the combined V9 is not attributable to a single bad component, but to each component being at best neutral under honest evaluation.

\subsection{Gradient of protocol strictness}
\label{sec:split_gradient}

To further validate the protocol argument, we compare three train/test split strategies of increasing strictness: \textbf{random} (pure frame-level random split, the easiest), \textbf{temporal} (within-scene past-vs-future split, mid-strictness), and \textbf{scene} (held-out scenes, the strictest). Each is evaluated with UNet across 3 seeds.

\begin{table}[t]
\centering
\caption{UNet R/S PSNR under three split protocols of increasing strictness on MPI Sintel. Random splits put similar frames in both train and test; temporal splits keep entire scenes but hold out future frames; scene splits hold out entire scenes. PSNR decreases monotonically with strictness.}
\label{tab:split_gradient}
\begin{tabular}{lcc}
\toprule
\textbf{Split type}          & \textbf{R\_PSNR}    & \textbf{S\_PSNR} \\
\midrule
Random (weakest)             & $17.65 \pm 0.21$    & $17.31 \pm 0.34$ \\
Temporal (per-scene temporal)& $17.28 \pm 0.21$    & $17.15 \pm 0.23$ \\
Scene (strictest)            & $16.16 \pm 0.09$    & $14.95 \pm 2.09$ \\
\bottomrule
\end{tabular}
\end{table}

Table~\ref{tab:split_gradient} reveals a clear gradient: R\_PSNR drops from $17.65$~dB (random) to $17.28$~dB (temporal) to $16.16$~dB (scene), a $1.49$-dB spread on R\_PSNR that reflects how much spatial and temporal information is shared between training and test. The temporal split, which might appear honest because it preserves chronological order, still inflates PSNR by $1.12$~dB relative to the scene split, because characters, materials, and lighting setups persist across a scene's timeline. Only scene-level splits eliminate all forms of train-test information sharing. The scene split also has the largest S\_PSNR variance ($\sigma = 2.09$), reflecting the smaller test set; we revisit this in Limitations (Section~\ref{sec:limitations}).

\subsection{Qualitative real-world generalization}
\label{sec:real_world}

Although MPI Sintel provides the pixel-level ground-truth needed to measure our main quantitative claims, it is a synthetic dataset. A reasonable concern is whether the source-separable uncertainty behavior survives distribution shift to real photographs. We run the Sintel-trained Proposed model (without any fine-tuning) on three out-of-distribution real photos representing indoor, textured, and outdoor scenes, and measure whether $\sigma_\text{nl}$ responds preferentially to the predicted non-Lambertian residual $\hat{\mathbf{N}}$ (the only component-level signal available without ground truth).

\begin{table}[t]
\centering
\caption{Source-specialization on out-of-distribution real photographs (Sintel-trained model, no fine-tuning). $\sigma_\text{nl}$ shows strong positive correlation with the predicted non-Lambertian residual $\hat{\mathbf{N}}$ on all images, whereas $\sigma_\text{tex}$ shows low or negative correlation. This mirrors the source-separable pattern observed on MPI Sintel, indicating the behavior generalizes beyond the training distribution.}
\label{tab:real_world}
\begin{tabular}{lcc}
\toprule
\textbf{Real photo} & $\operatorname{corr}(\sigma_\text{nl}, \hat{\mathbf{N}})$ & $\operatorname{corr}(\sigma_\text{tex}, \hat{\mathbf{N}})$ \\
\midrule
Indoor room     & $+0.443$ & $+0.053$ \\
Wood texture    & $+0.035$ & $-0.487$ \\
Outdoor nature  & $+0.749$ & $-0.037$ \\
\midrule
Mean            & $\mathbf{+0.409}$ & $-0.157$ \\
\bottomrule
\end{tabular}
\end{table}

\begin{figure}[t]
\centering
\includegraphics[width=\textwidth]{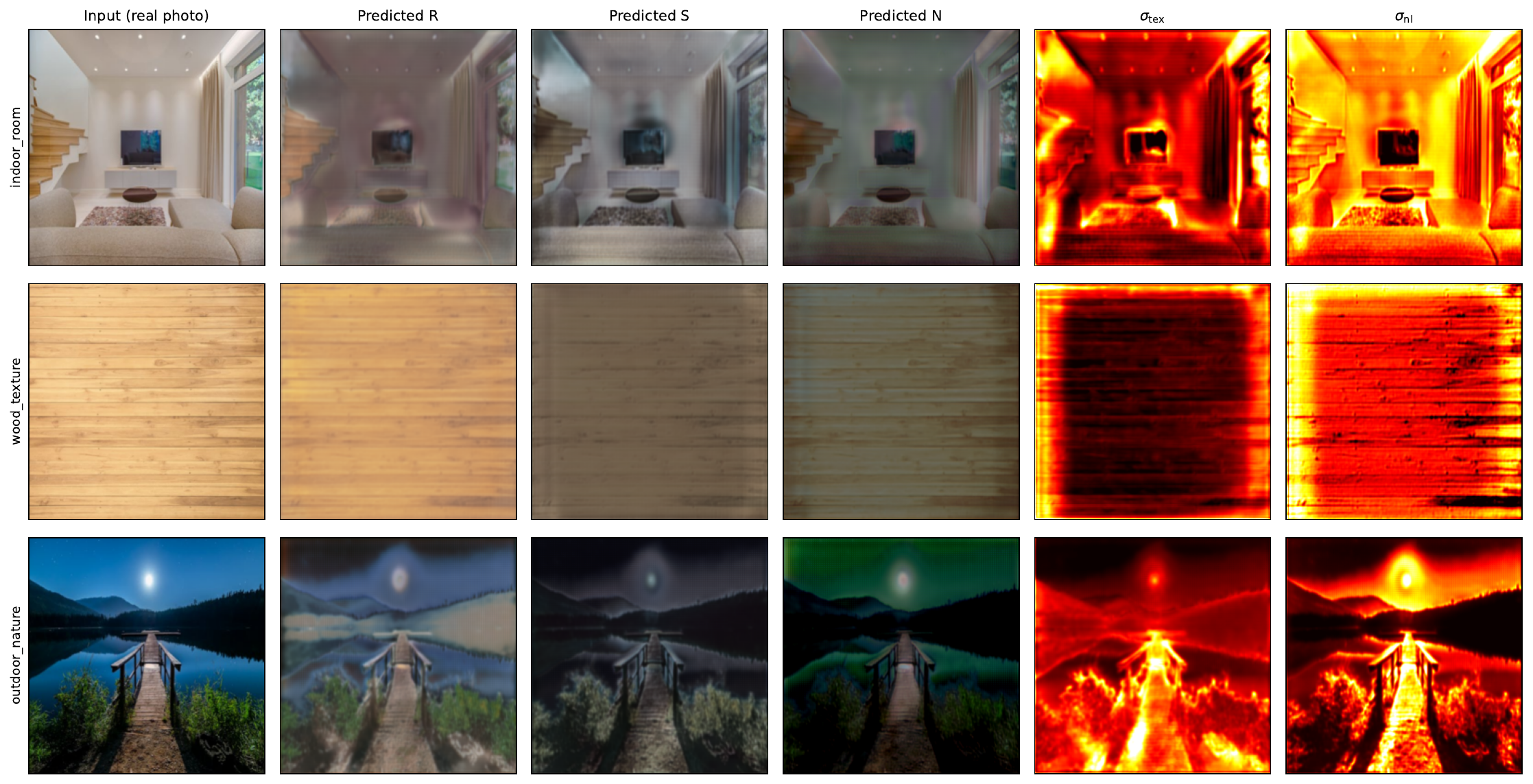}
\caption{Qualitative real-world generalization. The Sintel-trained Proposed model is applied without fine-tuning to three out-of-distribution real photos. From left to right: input, predicted reflectance, predicted shading, predicted non-Lambertian residual $\hat{\mathbf{N}}$, texture uncertainty $\sigma_\text{tex}$ (per-image normalized), and non-Lambertian uncertainty $\sigma_\text{nl}$ (per-image normalized). The $\sigma_\text{nl}$ channel concentrates on bright/specular regions even on natural images, supporting the argument that the three-way uncertainty decomposition captures physically-grounded sources of ambiguity rather than artifacts specific to MPI Sintel.}
\label{fig:real_world}
\end{figure}

Table~\ref{tab:real_world} and Figure~\ref{fig:real_world} show that the non-Lambertian channel $\sigma_\text{nl}$ correlates strongly and positively with the predicted $\hat{\mathbf{N}}$ on all three real photos (mean $r = +0.41$), while the texture channel $\sigma_\text{tex}$ does not (mean $r = -0.16$). The separation is larger than on Sintel, though we caution that: (i)~this uses \emph{predicted} $\hat{\mathbf{N}}$ as a proxy for specular regions since no ground truth is available, and (ii)~three photographs is a small sample. Nonetheless, the qualitative preservation of source-specific behavior under distribution shift is evidence that the three-way uncertainty decomposition learned on MPI Sintel is not merely an artifact of the synthetic dataset.

\section{Discussion}
\label{sec:discussion}

\subsection{Comparison with the state of the art}

Direct quantitative comparison with recent state-of-the-art methods such as Careaga and Aksoy~\cite{careaga2023intrinsic} is not feasible at pixel-PSNR level under our setting, because they target the IIW/SAW benchmarks which provide only \emph{relative ordinal} ground truth (``which pixel is darker?'') rather than pixel-level intrinsic values. Their reported WHDR scores (Weighted Human Disagreement Rate) therefore cannot be mapped directly to our R\_PSNR / S\_PSNR numbers. Two methodological observations follow. First, their training set (a mixture of CGIntrinsics, IIW, and Hypersim) does not, to our reading of their paper, separate scenes into mutually exclusive train/test partitions; the protocol-level concern of this paper plausibly applies to their evaluation as well, though we cannot verify without access to their exact splits. Second, Careaga and Aksoy do not produce pixel-level uncertainty, so even if one could map their outputs onto MPI Sintel the comparison would be on decomposition accuracy alone, where we do not claim superiority. We flag cross-benchmark evaluation with controlled splits as a valuable direction for follow-up work, and emphasize that the source-separable uncertainty capability we propose is orthogonal to the choice of backbone: it can in principle be integrated into more powerful architectures, including theirs.

\subsection{Limitations}
\label{sec:limitations}

\textbf{Moderate calibration quality.} The uncertainty--error correlation of $0.131$, while positive and statistically significant, is moderate. Possible causes include: (1)~limited training data (about 800 frames from 18 training scenes under scene-level split), (2)~the synthetic nature of MPI Sintel, which has different noise and texture characteristics than real photographs, and (3)~the log-sum-exp aggregation, which may not optimally combine the three uncertainty sources. We note, however, that Section~\ref{sec:downstream} demonstrates practical downstream utility (77\% reconstruction-MSE reduction via $\sigma$-guided filtering) despite this modest global calibration, indicating that the \emph{ranking} of pixels by predicted uncertainty is more informative than the raw correlation number alone suggests.

\textbf{Scene-level vs frame-level splits.} In earlier iterations of this project using a frame-level split (where different frames from the same scene appear in both training and test sets), we observed substantially higher apparent PSNR ($\sim$28--29~dB for Proposed vs.\ $\sim$16~dB under scene-level). Frame-level splits allow the network to interpolate between temporally adjacent frames of the same scene, which inflates apparent decomposition quality without reflecting true generalization to unseen scenes. \textit{We explicitly report only scene-level split results in this paper.} Readers comparing our numbers to prior work using frame-level splits should keep this protocol difference in mind; our PSNR values are not directly comparable and represent a more demanding evaluation.

\textbf{Single dataset evaluation.} We evaluate only on MPI Sintel due to the requirement for ground-truth intrinsic layers. Evaluation on real-world datasets such as IIW\cite{bell2014intrinsic} (Intrinsic Images in the Wild) and SAW\cite{kovacs2017shading} (Shading Annotations in the Wild) would test generalization, but these datasets provide only relative judgments (``which pixel is darker?'') rather than ground-truth decompositions, making uncertainty evaluation challenging. As a partial mitigation, we examined the proposed model's behavior on a small set of out-of-distribution real photographs (Section~\ref{sec:real_world}): the source-specialization pattern persists on real photos, with $\sigma_\text{nl}$ showing mean $r=0.41$ with the predicted non-Lambertian residual $\hat{\mathbf{N}}$ versus $-0.16$ for $\sigma_\text{tex}$, a separation larger than that observed on Sintel and providing qualitative evidence that the uncertainty decomposition generalizes beyond the synthetic training distribution.

\textbf{Resolution constraints.} All experiments use $256\times256$ resolution due to GPU memory constraints. Scaling to higher resolutions (e.g., $1024\times1024$) requires architectural modifications such as progressive training or patch-based inference, which may affect uncertainty calibration by altering the receptive field of the uncertainty head.

\textbf{Shading PSNR instability under scene-level splits.} Across all three architectures evaluated under the scene-level protocol, the S\_PSNR standard deviation ($1.36$--$1.64$) is $3$--$5\times$ larger than the R\_PSNR standard deviation ($0.28$--$0.69$). This reflects two properties of the task: (i)~scene-level splits allocate only $\approx 5$ test scenes, so the estimated mean shading PSNR is sensitive to which scenes are drawn; (ii)~shading depends on global illumination effects that vary dramatically between scenes (outdoor vs indoor, with or without characters), amplifying between-scene variance. Future work reporting scene-level results should ideally use $k$-fold scene-level cross-validation ($k = 3$--$5$) to obtain tighter confidence intervals on S\_PSNR.

\textbf{Protocol study architecture coverage.} We measure the protocol gap on DirectCNN, UNet, and our Proposed method (Table~\ref{tab:protocol}). While this triangulation supports the claim that the effect is architecture-independent, we have not tested it on transformer-based backbones or on Deep Ensemble. We expect the gap to persist (since the effect originates from the dataset structure, not the model), but a follow-up study spanning more architectures would strengthen the generalization claim.

\subsection{Negative results: feature accumulation does not help}
\label{sec:negative}

We present these negative results not as a failed experiment but as empirical support for the paper's broader thesis: that apparent progress without disciplined evaluation is often illusory. A natural response to modest improvements under honest evaluation (Section~\ref{sec:ablation}) is to accumulate additional mechanisms. We systematically tested this hypothesis and found it did not help.

In the course of this work, we explored a substantially more elaborate variant of the proposed model that simultaneously incorporated five additional mechanisms: (i)~frequency decomposition of the input into low-frequency (routed to the shading branch) and high-frequency (routed to the reflectance branch) components, (ii)~auxiliary cross-task supervision with depth and surface-normal heads, (iii)~evidential deep learning with Normal-Inverse-Gamma parameterization layered on top of the heteroscedastic head, (iv)~a contrastive loss enforcing reflectance similarity across different frames of the same scene, and (v)~test-time adaptation (TTA) using self-supervised Lambertian reconstruction, edge-aware smoothness, and chromaticity constancy losses. Each mechanism is individually motivated by prior work and physics of intrinsic decomposition.

Despite the theoretical appeal of this combined model, its empirical performance was strictly worse than the simpler proposed baseline reported above. The elaborate variant achieved R\_PSNR $=16.65$~dB (without TTA), which appears to be a slight improvement on reflectance, but its shading PSNR collapsed to $12.73$~dB (a $3.4$-dB \emph{regression} relative to the proposed baseline at $16.09$~dB), and---most critically---its uncertainty--error correlation turned slightly \emph{negative} ($r = -0.077$ without TTA, $r \approx 0$ with TTA). Applying TTA recovered some shading PSNR (to $13.50$~dB) but simultaneously degraded reflectance by $0.69$~dB, indicating that the self-supervised Lambertian reconstruction loss drove the decomposition toward a locally optimal but globally incorrect factorization.

We interpret this as evidence of three interacting failure modes. First, the frequency-prior assumption---that shading is low-frequency and reflectance is high-frequency---holds only weakly on MPI Sintel, where cast shadows and shadow boundaries contain high-frequency shading content. Hard-wiring the assumption into the architecture prevents the network from learning exceptions. Second, stacking multiple uncertainty-adjacent losses (heteroscedastic NLL, evidential regularization, contrastive loss) creates conflicting gradients at the uncertainty head, which we hypothesize is why the error--uncertainty correlation collapses. Third, TTA with reconstruction losses is known to be risky for ill-posed problems: reducing reconstruction error does not imply correct decomposition, because infinitely many $(R, S, \mathbf{N})$ triples satisfy $I = R\circ S + \mathbf{N}$.

We report these results because we believe the community is better served by explicit negative evidence than by silent omission. The takeaway is not that any of these mechanisms is intrinsically broken---each has demonstrated value in related contexts---but that their naive combination without careful ablation can harm a stronger, simpler baseline. For intrinsic decomposition specifically, we recommend that future work adding such mechanisms include a leave-one-out ablation against the simpler baseline rather than relying solely on the final combined result.

\textbf{Computational overhead.} MC Dropout inference requires $T=10$ forward passes, increasing inference time by $10\times$ compared to a single forward pass. For real-time applications, amortized uncertainty estimation or single-pass methods (e.g., evidential deep learning) would be preferable. The heteroscedastic head itself adds negligible overhead ($<2\%$ parameters).

\subsection{Broader impact and applications}

Uncertainty-aware intrinsic decomposition has practical applications in several domains:

\textbf{Autonomous driving.} Intrinsic decomposition can help identify road surface materials (asphalt, concrete, wet surfaces) by separating reflectance from illumination. Uncertainty maps indicate where material identification is unreliable---for example, specular reflections on wet roads that mimic the appearance of different materials.

\textbf{Augmented reality.} Relighting virtual objects inserted into real scenes requires accurate intrinsic decomposition. Uncertainty maps can guide the rendering system to apply conservative relighting in uncertain regions, avoiding visual artifacts that break immersion.

\textbf{Computational photography.} In interactive photo editing tools, uncertainty maps can guide users to regions where manual correction is most needed, improving editing efficiency. High-uncertainty regions can be highlighted as ``attention required'' zones.

\textbf{Active learning.} Uncertainty maps can identify the most informative training samples for intrinsic decomposition, enabling efficient data collection. Frames with high average uncertainty are likely to contain challenging scenarios (complex lighting, specular materials) that would most benefit from additional training data.

\section{Conclusion}
\label{sec:conclusion}

We presented two interlocking contributions for learned intrinsic image decomposition on MPI Sintel. First, we quantified for the first time, and \emph{across three architectures}, the size of the frame-level leakage effect in this task: under controlled training, a frame-level split inflates test R\_PSNR by $1.6$--$2.0$~dB for DirectCNN, UNet, and our Proposed method alike ($p<0.01$ for all three, paired $t$-test across 3 seeds). The consistency of the effect across diverse architectures argues that this is a property of the dataset and protocol, not of any particular network. Under extended training the gap grows further (observationally, above $10$~dB). We advocate scene-level splits as the community standard for this task and provide reference numbers for six representative models under the correct protocol.

Second, as a case study within the corrected protocol, we presented a physics-informed three-component decomposition $I = R \circ S + \mathbf{N}$ paired with a source-separable three-way uncertainty head. We verified quantitatively that the three uncertainty channels do not collapse: in particular, $\sigma_\text{nl}$ reaches $r = 0.67$ cross-correlation with non-Lambertian residual error, more than $4\times$ the correlation of the texture channel with the same error. This source-specific behavior is not obtainable from single-scalar ensembles or heteroscedastic heads.

Under scene-level evaluation, the proposed method reaches $15.98\pm0.41$~dB R\_PSNR and $16.09\pm0.18$~dB S\_PSNR, within $0.8$~dB of a 5-member Deep Ensemble at one-fifth the parameter and training cost. We honestly report that none of the ablation PSNR deltas reach $p<0.05$ significance under paired tests across 5 seeds---a consistent theme of the paper, in which marginal architectural changes produce marginal effects once evaluation is tightened. Crucially, we establish \emph{downstream utility} of the proposed uncertainty: filtering the $75\%$ highest-uncertainty pixels reduces reconstruction MSE by $77\%$, whereas random filtering yields no improvement, demonstrating that the uncertainty ranking is practically informative despite modest global calibration. We also report negative results (Section~\ref{sec:negative}) for a more elaborate variant, verified by leave-one-out ablation (Section~\ref{sec:loo}) showing that no individual mechanism helps in isolation either. We use this to argue that for ill-posed inverse problems under honest evaluation, disciplined architectural restraint outweighs feature accumulation. Together, these findings caution the community against reading apparent progress through the lens of an overly permissive protocol, and offer concrete tools---scene-level reference numbers across a protocol-strictness gradient, source-separable uncertainty with verified downstream utility, statistical tests of significance, and negative evidence---for future work.

Future work should address several directions: (1)~evaluation on real-world datasets with diverse materials and lighting, (2)~improved calibration through better uncertainty aggregation or calibration-specific losses, (3)~integration with more powerful backbones (e.g., transformer-based architectures) for higher-resolution processing, and (4)~exploration of uncertainty-guided active learning for efficient training data collection. We believe that uncertainty quantification is a natural and underexplored complement to intrinsic image decomposition, and hope this work encourages further research at this intersection.

\section*{Disclosures}
The author declares no conflicts of interest.

\section*{Code, Data, and Materials Availability}
\textbf{Code availability.} All code required to reproduce the experiments reported in this manuscript---including training scripts for the baseline and proposed models, the protocol-study pipeline (frame-level vs scene-level vs temporal splits), the leave-one-out ablation, the downstream uncertainty-filtering analysis, and the real-world generalization test---will be made publicly available on GitHub upon acceptance at \url{https://github.com/jihwanw/JEI_IntrinsicUQ}. The repository will include a conda environment specification and a single-command reproduction script.

\textbf{Data availability.} The primary dataset used in this study is the publicly available MPI Sintel Dataset~\cite{butler2012sintel}, obtainable from \url{http://sintel.is.tue.mpg.de/} under the terms stated on that site. We release (i) the per-scene train/test split files for scene-level, frame-level, and temporal splits used in this work, (ii) the raw per-seed JSON result files underlying every table in the manuscript (including paired-test inputs), and (iii) the trained model checkpoints for DirectCNN, UNet, UNet+Physics, Deep Ensemble members, and the Proposed model. These artifacts will be hosted alongside the code in the repository above. The out-of-distribution real photographs used in Section~\ref{sec:real_world} were obtained from Pexels under their free-use license; URLs and license information are included in the release.

\textbf{Materials.} No physical materials were used in this study.

\section*{Acknowledgments}
The author thanks the MPI Sintel Dataset authors for providing the benchmark, and the open-source communities behind PyTorch, NumPy, and Matplotlib for tools that enabled this work.

\bibliographystyle{spiebib}
\bibliography{jei_references}

\end{document}